\documentclass[submission,copyright,creativecommons]{eptcs}
\providecommand{\event}{TTC 2011} 

\usepackage{multirow}
\usepackage{graphicx}
\usepackage{subfigure}
\usepackage{url}
\usepackage{amssymb}
\usepackage{xspace}

\usepackage{amsmath}
\usepackage{listings}

\makeindex

\newcommand{\eg}{e.\,g.\ }

\lstdefinelanguage{groovy} {
    emph={println, new, tokenize, each, def, static, for, if, else, in, assert},
    emphstyle=\bfseries,
    morecomment=[l]{//},
    basicstyle=\fontfamily{pcr}\scriptsize,
    string=[b]",
    showstringspaces=false
}

\newcommand{\operation}[1]{\textsf{#1}}

\newcommand{\myparagraph}[1]{\vspace{.5em}\par\noindent\textbf{#1}}

\title{Solving the TTC 2011 Reengineering Case with Edapt}
\author{Markus Herrmannsdoerfer
\institute{Institut f\"ur Informatik,
  Technische Universit\"{a}t M\"{u}nchen 
}
\email{herrmama@in.tum.de}
}
\def\titlerunning{Solving the TTC 2011 Reengineering Case with Edapt}
\def\authorrunning{M. Herrmannsdoerfer}
\begin{document}
\maketitle

\begin{abstract}
This paper gives an overview of the Edapt solution to the reengineering case \cite{programunderstandingcase} of the Transformation Tool Contest 2011.
\end{abstract}

\section{Edapt in a Nutshell}

Edapt\footnote{\url{http://www.eclipse.org/edapt}} is a transformation tool tailored for the migration of models in response to metamodel adaptation.
Edapt is an official Eclipse tool derived from the research prototype COPE.

\begin{figure}[htb]
\centering
		\includegraphics[scale=0.6]{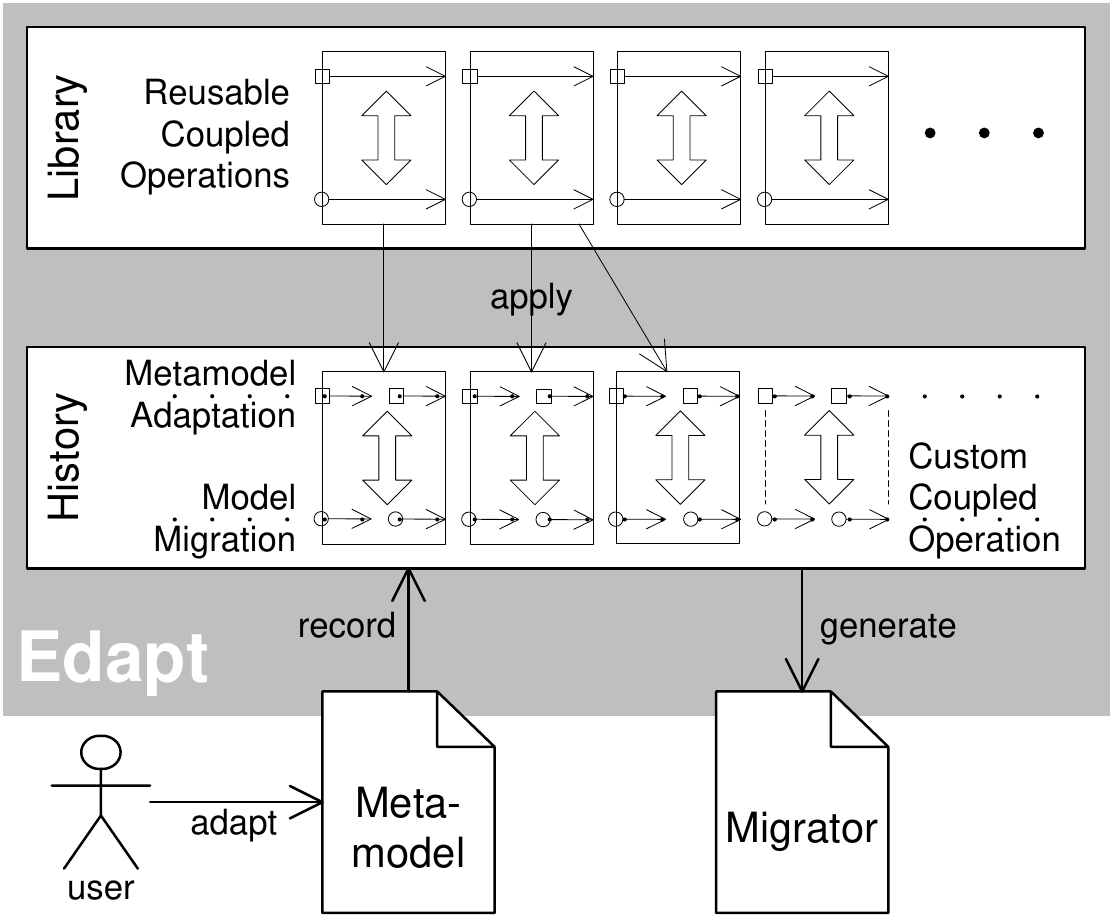}
				\vskip -5pt
\caption{Overview of Edapt}
\label{fig:overview}
\end{figure}

\myparagraph{Modeling the Coupled Evolution.}
As depicted by Figure~\ref{fig:overview}, Edapt specifies the me\-ta\-mo\-del adaptation as a sequence of operations in an explicit history model.
The operations can be enriched with instructions for model migration to form so-called coupled operations.
Edapt provides two kinds of coupled operations according to the automatability of the model migration~\cite{Herrmannsdoerfer2009_COPE-AutomatingCoupledEvolutionofMetamodelsandModels}: reusable and custom coupled operations.

Reuse of recurring migration specifications allows to reduce the effort associated with building a model migration~\cite{Herrmannsdoerfer2008_AutomatabilityofCoupledEvolutionofMetamodelsandModelsinPractice}.
Edapt thus provides \emph{reusable coupled operations} which make metamodel adaptation and model migration independent of the specific metamodel through parameters and constraints restricting the applicability of the operation.
An example for a reusable coupled operation is \emph{Enumeration to Sub Classes} which replaces an enumeration attribute with subclasses for each literal of the enumeration.
Currently, Edapt comes with a library of over 60 reusable coupled operations~\cite{Herrmannsdoerfer2010_AnExtensiveCatalogofOperatorsfortheCoupledEvolutionofMetamodelsandModels}.
By means of studying real-life metamodel histories, we have shown that, in practice, most of the coupled evolution can be covered by reusable coupled operations~\cite{Herrmannsdoerfer2008_AutomatabilityofCoupledEvolutionofMetamodelsandModelsinPractice,Herrmannsdoerfer2010_LanguageEvolutioninPracticeTheHistoryofGMF}.

Migration specifications can become so specific to a certain metamodel that reuse does not make sen\-se~\cite{Herrmannsdoerfer2008_AutomatabilityofCoupledEvolutionofMetamodelsandModelsinPractice}.
To express these complex migrations, Edapt allows the user to define a custom coupled operation by manually encoding a model migration for a metamodel adaptation in a Turing-complete language~\cite{Herrmannsdoerfer2008_COPEALanguagefortheCoupledEvolutionofMetamodelsandModels}.
By softening the conformance of the model to the metamodel within a coupled operation, both metamodel adaptation and model migration can be specified as in-place transformations, requiring only to specify the difference.
A transaction mechanism ensures conformance at the boundaries of the coupled operation.

\begin{figure}[!tb]
\centering
		\includegraphics[width=.85\textwidth]{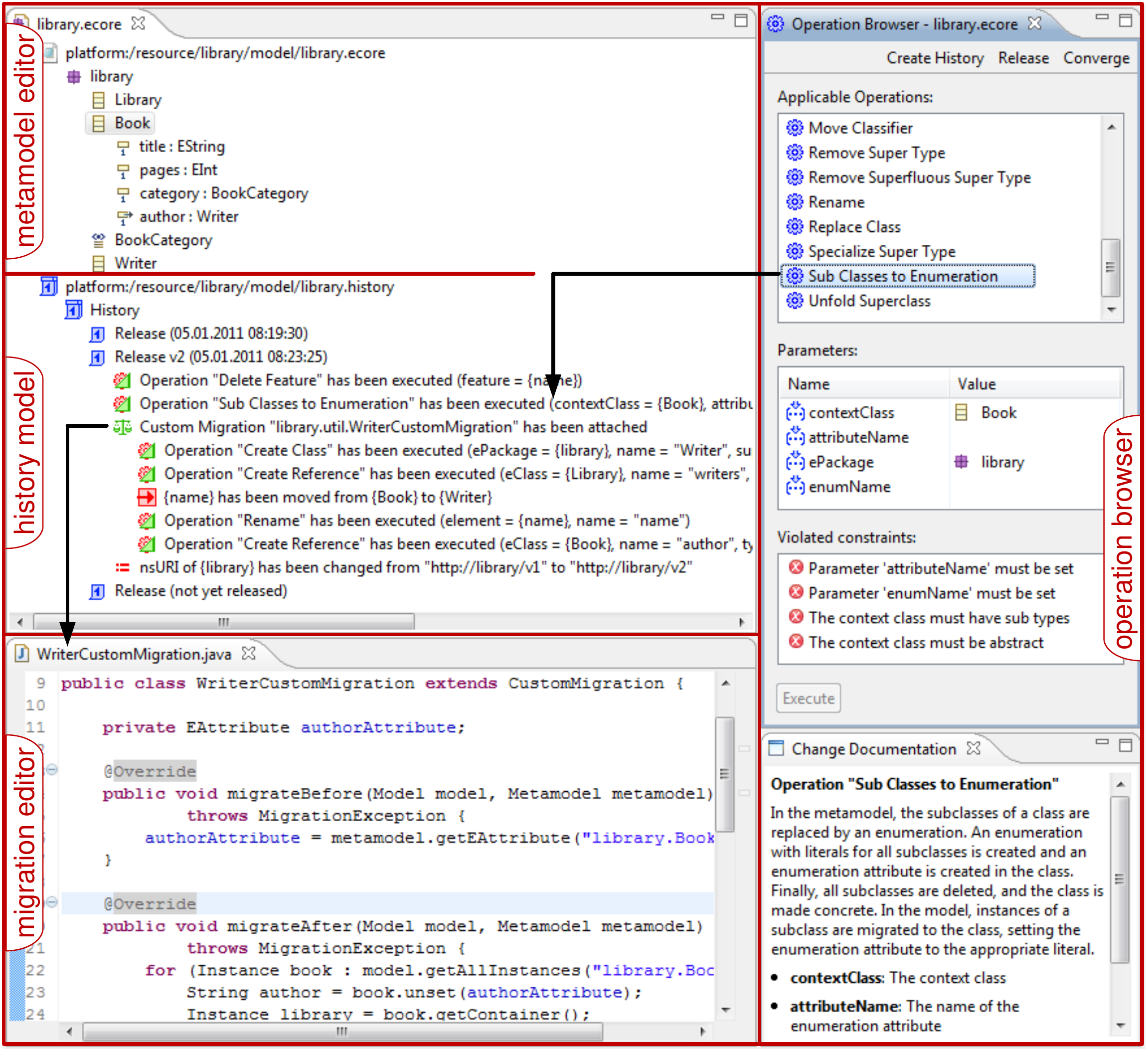}
		\vskip -5pt
\caption{User interface of Edapt}
\label{fig:screenshot}
\end{figure}

\myparagraph{Recording the Coupled Evolution.}
To not lose the intention behind the metamodel adaptation, Edapt is intended to be used already when adapting the metamodel.
Therefore, Edapt's user interface, which is depicted in Figure~\ref{fig:screenshot}, is directly integrated into the existing EMF \emph{metamodel editor}.
The user interface provides access to the \emph{history model} in which Edapt records the sequence of coupled operations.
An initial history can be created for an existing metamodel by invoking \emph{Create History} in the \emph{operation browser} which also allows the user to \emph{Release} the metamodel.

The user can adapt the metamodel by applying reusable coupled operations through the \emph{operation browser}.
The operation browser allows to set the parameters of a reusable coupled operation, and gives feedback on the operation's applicability based on the constraints.
When a reusable coupled operation is executed, its application is automatically recorded in the history model.
Figure~\ref{fig:screenshot} shows the operation \operation{Sub Classes to Enumeration} being selected in the operation browser and recorded to the history model.

The user needs to perform a custom coupled operation only, in case no reusable coupled operation is available for the change at hand.
First, the metamodel is directly adapted in the metamodel editor, in response to which the changes are automatically recorded in the history.
A migration can later be attached to the sequence of metamodel changes.
Figure~\ref{fig:screenshot} shows the \emph{migration editor} to encode the custom migration in Java.

\section{Reengineering Case}

\begin{figure}
	\centering
		\includegraphics[scale=0.6]{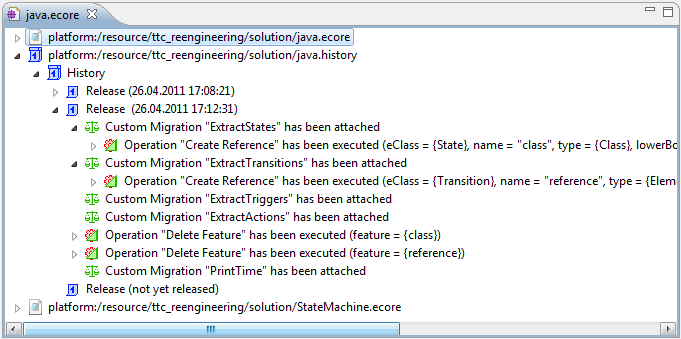}
		\caption{History model for the reengineering case}
	\label{fig:reengineering}
\end{figure}

Figure~\ref{fig:reengineering} shows the history model that solves the reengineering case.
The history model was initialized for both the Java and the statemachine metamodel.
The history model is modularized into custom coupled operations for extracting states, transitions, triggers and actions from the Java model.
In the first steps, trace associations from the statemachine to the Java metamodel are added, which are deleted at the end by reusable coupled operations.
In the following, we briefly explain the different steps.
The complete solution is available in the appendix, through a SHARE demo \cite{programunderstandingcase} and in the repository of the Eclipse Edapt project\footnote{\url{http://dev.eclipse.org/svnroot/modeling/org.eclipse.emft.edapt/trunk/examples/ttc_reengineering}}.

\myparagraph{Core Task.}
To extract the states from the Java model (see Listing~\ref{sec:appendix_ExtractStates}), we first create the statemachine in a separate model resource (lines 27-43).
Then we search for the abstract base class that denotes states in the Java model (lines 58-66).
From this class, we navigate opposite to inheritance associations to obtain all non-abstract subclasses (lines 68-103).
In Edapt, opposite navigation can be performed very efficiently using the helper method \textsf{getInverse} (\eg lines 83/84).
Finally, we create a state for each non-abstract subclass (lines 45-56).
We also associate a state to its implementing class through the association \textsf{class} created for this purpose (line 53).

To extract transitions from the Java model (see Listing~\ref{sec:appendix_ExtractTransitions}), for each class implementing a target state, we search for \textsf{ElementReference}s to it in the classes implementing source states (lines 18-39).
This can again be performed very efficiently using the inverse navigation of Edapt.
We also have to check whether the \textsf{ElementReference}s are part of a call to the \textsf{activate} method before creating a transition (lines 41-57).
We use the association \textsf{class} created in the last custom coupled operation to navigate from a state in the statemachine model to its implementing class in the Java model (\eg line 22), and vice versa (\eg lines 29/30).
We associate a transition to its implementing \textsf{ElementReference} through the association \textsf{reference} created for this purpose (lines 60-69).

\myparagraph{Extension 1: Triggers.}
We implemented the logic as explained by the case description to extract triggers from the Java model for each transition in the statemachine model (see Listing~\ref{sec:appendix_ExtractTriggers}).
We use the association \textsf{reference} to navigate from a transition in the statemachine model to its implementing \textsf{ElementReference} in the Java model.

\myparagraph{Extension 2: Actions.}
We implemented the logic as explained by the case description to extract actions from the Java model for each transition in the statemachine model (see Listing~\ref{sec:appendix_ExtractActions}).
Again, we use the association \textsf{reference} to navigate between transitions and their implementations.

\section{Conclusion}

We discuss the solution with respect to the evaluation criteria defined in the case description.

\myparagraph{Understandability.}
We use Edapt's mechanism to compose coupled operations to modularize the solution into a number of understandable steps.
Moreover, each of the custom migrations can again be split up into a number of methods to increase understandability.
However, the implementation in Java makes the migration slightly more difficult to understand than a specialized DSL.

\myparagraph{Conciseness.}
The solution is quite concise, since we can rely on Java's inheritance mechanism to reuse recurring methods---\eg \textsf{getContainerOfType} in the super class \textsf{ReengineeringCustomMigration} to obtain the direct or indirect parent of a certain type.
Moreover, the API with which migrations are implemented provides methods that foster a concise implementation---\eg methods that resolve elements from the metamodel by their fully qualified name.
Even though the implementation in Java is quite concise and understandable, a specialized DSL could further improve conciseness and understandability.
However, we can rely on the strong Java tooling to implement, refactor and debug the solution.

\myparagraph{Completeness.}
We believe that the solution is complete, since it covers all cases specified in the case description.
Moreover, we obtain the same statemachine model as the reference models for the test models.

\myparagraph{Performance.}
Due to the efficient inverse navigation of Edapt which has the same performance as the forward navigation, the transformation has a good performance on the test models.
Since all Java test models contain the same statemachine model, the transformation itself takes the same time on all test models: 0.5 to 1~s on the SHARE virtual machine.
However, loading the models takes longer for the bigger ones.


\bibliographystyle{eptcs}
\bibliography{literature}

\begin{thebibliography}{1}
\providecommand{\bibitemdeclare}[2]{}
\providecommand{\urlprefix}{Available at }
\providecommand{\url}[1]{\texttt{#1}}
\providecommand{\href}[2]{\texttt{#2}}
\providecommand{\urlalt}[2]{\href{#1}{#2}}
\providecommand{\doi}[1]{doi:\urlalt{http://dx.doi.org/#1}{#1}}
\providecommand{\bibinfo}[2]{#2}

\bibitemdeclare{inproceedings}{Herrmannsdoerfer2008_AutomatabilityofCoupledEvo%
lutionofMetamodelsandModelsinPractice}
\bibitem{Herrmannsdoerfer2008_AutomatabilityofCoupledEvolutionofMetamodelsandM%
odelsinPractice}
\bibinfo{author}{Markus Herrmannsdoerfer}, \bibinfo{author}{Sebastian Benz} \&
  \bibinfo{author}{Elmar Juergens} (\bibinfo{year}{2008}):
  \emph{\bibinfo{title}{Automatability of Coupled Evolution of Metamodels and
  Models in Practice}}.
\newblock In: {\sl \bibinfo{booktitle}{MoDELS '08}},
  \doi{10.1007/978-3-540-87875-9\_45}.

\bibitemdeclare{inproceedings}{Herrmannsdoerfer2008_COPEALanguagefortheCoupled%
EvolutionofMetamodelsandModels}
\bibitem{Herrmannsdoerfer2008_COPEALanguagefortheCoupledEvolutionofMetamodelsa%
ndModels}
\bibinfo{author}{Markus Herrmannsdoerfer}, \bibinfo{author}{Sebastian Benz} \&
  \bibinfo{author}{Elmar Juergens} (\bibinfo{year}{2008}):
  \emph{\bibinfo{title}{{COPE}: A Language for the Coupled Evolution of
  Metamodels and Models}}.
\newblock In: {\sl \bibinfo{booktitle}{MCCM '08}}.

\bibitemdeclare{inproceedings}{Herrmannsdoerfer2009_COPE-AutomatingCoupledEvol%
utionofMetamodelsandModels}
\bibitem{Herrmannsdoerfer2009_COPE-AutomatingCoupledEvolutionofMetamodelsandMo%
dels}
\bibinfo{author}{Markus Herrmannsdoerfer}, \bibinfo{author}{Sebastian Benz} \&
  \bibinfo{author}{Elmar Juergens} (\bibinfo{year}{2009}):
  \emph{\bibinfo{title}{{COPE} - Automating Coupled Evolution of Metamodels and
  Models}}.
\newblock In: {\sl \bibinfo{booktitle}{{ECOOP} '09}},
  \doi{10.1007/978-3-642-03013-0\_4}.

\bibitemdeclare{inproceedings}{Herrmannsdoerfer2010_LanguageEvolutioninPractic%
eTheHistoryofGMF}
\bibitem{Herrmannsdoerfer2010_LanguageEvolutioninPracticeTheHistoryofGMF}
\bibinfo{author}{Markus Herrmannsdoerfer}, \bibinfo{author}{Daniel Ratiu} \&
  \bibinfo{author}{Guido Wachsmuth} (\bibinfo{year}{2009}):
  \emph{\bibinfo{title}{Language Evolution in Practice: The History of {GMF}}}.
\newblock In: {\sl \bibinfo{booktitle}{SLE '09}},
  \doi{10.1007/978-3-642-12107-4\_3}.

\bibitemdeclare{inproceedings}{Herrmannsdoerfer2010_AnExtensiveCatalogofOperat%
orsfortheCoupledEvolutionofMetamodelsandModels}
\bibitem{Herrmannsdoerfer2010_AnExtensiveCatalogofOperatorsfortheCoupledEvolut%
ionofMetamodelsandModels}
\bibinfo{author}{Markus Herrmannsdoerfer}, \bibinfo{author}{Sander Vermolen} \&
  \bibinfo{author}{Guido Wachsmuth} (\bibinfo{year}{2010}):
  \emph{\bibinfo{title}{An Extensive Catalog of Operators for the Coupled
  Evolution of Metamodels and Models}}.
\newblock In: {\sl \bibinfo{booktitle}{SLE '10}},
  \doi{10.1007/978-3-642-19440-5\_10}.

\bibitemdeclare{inproceedings}{programunderstandingcase}
\bibitem{programunderstandingcase}
\bibinfo{author}{Tassilo Horn} (\bibinfo{year}{2011}):
  \emph{\bibinfo{title}{Program Understanding: A Reengineering Case for the
  Transformation Tool Contest}}.
\newblock In \bibinfo{editor}{Pieter {Van Gorp}}, \bibinfo{editor}{Steffen
  Mazanek} \& \bibinfo{editor}{Louis Rose}, editors: {\sl
  \bibinfo{booktitle}{{TTC} 2011: Fifth Transformation Tool Contest, Z\"urich,
  Switzerland, June 29-30 2011}}, \bibinfo{publisher}{{EPTCS}}.

\end{thebibliography}

\newpage
\appendix

\section{Solution}

\subsection{Base Class for Custom Migrations of the Reengineering Case}
\begin{lstlisting}[language=java]
import org.eclipse.emf.edapt.migration.CustomMigration;
import org.eclipse.emf.edapt.migration.Instance;


public abstract class ReengineeringCustomMigration extends CustomMigration {

	/**
	 * Get the direct or indirect container of an element that is instance of a
	 * certain type.
	 */
	protected Instance getContainerOfType(Instance instance, String type) {
		Instance container = instance;
		while (container != null && !container.instanceOf(type)) {
			container = container.getContainer();
		}
		return container;
	}
}
\end{lstlisting}
\vspace{1em}

\subsection{Extract States}
\label{sec:appendix_ExtractStates}
\begin{lstlisting}[language=java]
import java.util.ArrayList;
import java.util.List;

import org.eclipse.emf.common.util.URI;
import org.eclipse.emf.edapt.migration.CustomMigration;
import org.eclipse.emf.edapt.migration.Instance;
import org.eclipse.emf.edapt.migration.Metamodel;
import org.eclipse.emf.edapt.migration.MigrationException;
import org.eclipse.emf.edapt.migration.Model;
import org.eclipse.emf.edapt.migration.ModelResource;

public class ExtractStates extends CustomMigration {
	
	@Override
	public void migrateBefore(Model model, Metamodel metamodel)
			throws MigrationException {
		PrintTime.start = System.currentTimeMillis();
	}

	@Override
	public void migrateAfter(Model model, Metamodel metamodel)
			throws MigrationException {
		Instance stateMachine = createStateMachine(model);
		createStates(model, stateMachine);
	}

	/** Create the state machine. */
	private Instance createStateMachine(Model model) {
		ModelResource resource = createResultResource(model);
		Instance stateMachine = model.newInstance("statemachine.StateMachine");
		resource.getRootInstances().add(stateMachine);
		return stateMachine;
	}

	/** Create the resource to store the state machine in. */
	private ModelResource createResultResource(Model model) {
		URI uri = model.getResources().get(0).getUri();
		URI resultUri = uri
				.trimSegments(1)
				.appendSegment(uri.trimFileExtension().lastSegment() + "result")
				.appendFileExtension(uri.fileExtension());
		return model.newResource(resultUri);
	}

	/** Create the states. */
	private void createStates(Model model, Instance stateMachine) {
		Instance stateClass = getStateClass(model);
		for (Instance subClass : getAllSubClasses(stateClass)) {
			if (!isAbstract(subClass)) {
				Instance state = model.newInstance("statemachine.State");
				state.set("name", subClass.get("name"));
				stateMachine.add("states", state);
				state.set("class", subClass);
			}
		}
	}

	/** Get the class that is super class of all classes representing states. */
	private Instance getStateClass(Model model) {
		for (Instance c : model.getAllInstances("classifiers.Class")) {
			if ("State".equals(c.get("name"))) {
				return c;
			}
		}
		return null;
	}

	/** Get all direct and indirect sub classes of a class. */
	private List<Instance> getAllSubClasses(Instance c) {
		List<Instance> subTypes = new ArrayList<Instance>();
		for (Instance subType : getSubClasses(c)) {
			if (!subTypes.contains(subType)) {
				subTypes.add(subType);
			}
			subTypes.addAll(getAllSubClasses(subType));
		}
		return subTypes;
	}

	/** Get all direct sub classes of a class. */
	private List<Instance> getSubClasses(Instance c) {
		List<Instance> subClasses = new ArrayList<Instance>();
		List<Instance> classifierReferences = c
				.getInverse("types.ClassifierReference.target");
		for (Instance classifierReference : classifierReferences) {
			Instance container = classifierReference.getContainer()
					.getContainer();
			if (container.instanceOf("classifiers.Class")) {
				subClasses.add(container);
			}
		}
		return subClasses;
	}

	/** Check whether a class is abstract. */
	private boolean isAbstract(Instance c) {
		for (Instance modifier : c.getLinks("annotationsAndModifiers")) {
			if (modifier.instanceOf("modifiers.Abstract")) {
				return true;
			}
		}
		return false;
	}
}
\end{lstlisting}
\vspace{1em}

\subsection{Extract Transitions}
\label{sec:appendix_ExtractTransitions}
\begin{lstlisting}[language=java]
import java.util.List;

import org.eclipse.emf.edapt.migration.Instance;
import org.eclipse.emf.edapt.migration.Metamodel;
import org.eclipse.emf.edapt.migration.MigrationException;
import org.eclipse.emf.edapt.migration.Model;

public class ExtractTransitions extends ReengineeringCustomMigration {

	@Override
	public void migrateAfter(Model model, Metamodel metamodel)
			throws MigrationException {
		Instance stateMachine = model.getInstances("statemachine.StateMachine")
				.get(0);
		createTransitions(model, stateMachine);
	}

	/** Create the transitions. */
	private void createTransitions(Model model, Instance stateMachine) {
		List<Instance> states = stateMachine.getLinks("states");
		for (Instance targetState : states) {
			Instance targetClass = targetState.getLink("class");
			List<Instance> references = targetClass
					.getInverse("references.ElementReference.target");
			for (Instance reference : references) {
				if (isTransition(reference)) {
					Instance sourceClass = getContainerOfType(reference,
							"classifiers.Class");
					List<Instance> sourceStates = sourceClass
							.getInverse("statemachine.State.class");
					if (!sourceStates.isEmpty()) {
						Instance sourceState = sourceStates.get(0);
						createTransition(model, stateMachine, sourceState,
								targetState, reference);
					}
				}
			}
		}
	}

	/**
	 * Check whether a certain reference to a state class represents a
	 * transition.
	 */
	private boolean isTransition(Instance reference) {
		Instance next = reference.getLink("next");
		if (next != null && next.instanceOf("references.MethodCall")) {
			if ("Instance".equals(next.getLink("target").get("name"))) {
				next = next.get("next");
				if (next != null && next.instanceOf("references.MethodCall")) {
					if ("activate".equals(next.getLink("target").get("name"))) {
						return true;
					}
				}
			}
		}
		return false;
	}

	/** Create a transition from a source to the target state. */
	private void createTransition(Model model, Instance stateMachine,
			Instance sourceState, Instance targetState, Instance reference) {
		Instance transition = model.newInstance("statemachine.Transition");
		transition.set("src", sourceState);
		transition.set("dst", targetState);
		transition.set("reference", reference);
		stateMachine.add("transitions", transition);
	}
}
\end{lstlisting}
\vspace{1em}

\subsection{Extract Triggers}
\label{sec:appendix_ExtractTriggers}
\begin{lstlisting}[language=java]
import org.eclipse.emf.edapt.migration.Instance;
import org.eclipse.emf.edapt.migration.Metamodel;
import org.eclipse.emf.edapt.migration.MigrationException;
import org.eclipse.emf.edapt.migration.Model;

public class ExtractTriggers extends ReengineeringCustomMigration {

	@Override
	public void migrateAfter(Model model, Metamodel metamodel)
			throws MigrationException {
		for (Instance transition : model
				.getAllInstances("statemachine.Transition")) {
			Instance reference = transition.get("reference");
			transition.set("trigger", getTrigger(reference));
		}
	}

	/** Get the trigger of a transition. */
	private String getTrigger(Instance reference) {
		Instance method = getContainerOfType(reference, "members.ClassMethod");
		String methodName = method.get("name");
		if ("run".equals(methodName)) {
			Instance switchCase = getContainerOfType(reference,
					"statements.NormalSwitchCase");
			if (switchCase != null) {
				return switchCase.getLink("condition").getLink("target")
						.get("name");
			}
			Instance catchBlock = getContainerOfType(reference,
					"statements.CatchBlock");
			if (catchBlock != null) {
				return catchBlock.getLink("parameter").getLink("typeReference")
						.getLinks("classifierReferences").get(0)
						.getLink("target").get("name");
			}
		} else {
			return methodName;
		}
		return "--";
	}
}
\end{lstlisting}
\vspace{1em}

\subsection{Extract Actions}
\label{sec:appendix_ExtractActions}
\begin{lstlisting}[language=java]
import org.eclipse.emf.edapt.migration.Instance;
import org.eclipse.emf.edapt.migration.Metamodel;
import org.eclipse.emf.edapt.migration.MigrationException;
import org.eclipse.emf.edapt.migration.Model;

public class ExtractActions extends ReengineeringCustomMigration {

	@Override
	public void migrateAfter(Model model, Metamodel metamodel)
			throws MigrationException {
		for (Instance transition : model
				.getAllInstances("statemachine.Transition")) {
			Instance reference = transition.get("reference");
			transition.set("action", getAction(reference));
		}
	}

	/** Get the action of a transition. */
	private String getAction(Instance reference) {
		Instance container = getContainerOfType(reference,
				"statements.StatementListContainer");
		for (Instance statement : container.getLinks("statements")) {
			if (statement.instanceOf("statements.ExpressionStatement")) {
				Instance expression = statement.getLink("expression");
				if (expression.instanceOf("references.MethodCall")) {
					if ("send".equals(expression.getLink("target").get("name"))) {
						return expression.getLinks("arguments").get(0)
								.getLink("next").getLink("target").get("name");
					}
				}
			}
		}
		return "--";
	}
}
\end{lstlisting}
\vspace{1em}

\subsection{Print Time}
\begin{lstlisting}[language=java]
import org.eclipse.emf.edapt.migration.CustomMigration;
import org.eclipse.emf.edapt.migration.Metamodel;
import org.eclipse.emf.edapt.migration.MigrationException;
import org.eclipse.emf.edapt.migration.Model;

public class PrintTime extends CustomMigration {

	public static long start;

	@Override
	public void migrateAfter(Model model, Metamodel metamodel)
			throws MigrationException {
		long time = System.currentTimeMillis() - start;
		System.out.println(time + " ms");
	}
}
\end{lstlisting}
\vspace{1em}

\end{document}